

\input phyzzx


\newdimen\fullhsize
\newbox\leftcolumn
\def\Full{n }

\message{ Reduce to half size? [yn] }
\read-1 to \Size
\ifx\Size\Full	
\else		


\let\seventeenbf=\fourteenbf
\let\fourteenbf=\twelvebf

\tenpoint

\normalbaselineskip=16pt plus 2pt minus 1pt
\baselineskip=16pt plus 2pt minus 1pt



\tolerance=1000
\fullhsize=10truein \hsize=4.75truein
\def\fulline{\hbox to \fullhsize}
\vsize=7truein
\hfuzz=5pt
\hoffset=-.56truein	\voffset=-.4truein

\let\lr=L 

\output={\if L\lr
   \global\setbox\leftcolumn=\columnbox \global\let\lr=R
   \else\doubleformat \global\let\lr=L\fi
   \advancepageno
   \ifnum\outputpenalty>-20000 \else\dosupereject\fi}

\def\doubleformat{\shipout\vbox{\makeheadline
   \fulline{\box\leftcolumn\hfil\columnbox} }}

\def\columnbox{\leftline{\vbox{{\pagebody\makefootline}}}}

\fi



\hfuzz=10pt

\def\A{{\cal A}}
\def\D{{\rm d}}
\def\PHI{\partial_x\zeta}
\def\PHIDOT{\dot\zeta}
\def\PI{\pi_\zeta}

\def\Exp#1{{\rm e}^{#1}}

\def\Chapter#1{\sectionnumber=0 {\bf\chapter{#1}}}
\def\refmark#1{[#1]}		
\def\scrunch{ \multiply\baselineskip by 3 \divide\baselineskip by 4}
\def\ie{{\it i.e.}}

\def\AP{{\it Ann.\ Phys.\ }}

\def\MPL{{\it Mod.\ Phys.\ Lett.\ }}

\def\NP{{\it Nucl.\ Phys.\ }}
\def\PL{{\it Phys.\ Lett.\ }}
\def\PR{{\it Phys.\ Rev.\ }}

\def\art#1{}
\def\art#1{{\sl ``#1''}}


\REF\klebanov{I.~Klebanov,
   \art{String theory in two dimensions,}
   Lectures at ICTP Spring School on String Theory and Quantum Gravity,
   Trieste, April 1991, Princeton preprint PUPT-1271, hep-th/9108019}

\REF\shenker{S.H.~Shenker,
   \art{The strength of non-perturbative effects in string theory,}
   Rutgers preprint RU-90-47, 1990, presented at the Carg\`ese Workshop on
   Random Surfaces, Quantum Gravity and Strings}

\REF\jevicki{A.~Jevicki,
   \art{Non-perturbative collective field theory,}
   \NP {\bf B376} (1992) 75}
\REF\MPR{G.~Moore, M.R.~Plesser and S. Ramgoolam,
   \art{Exact $S$-matrix for 2D string theory,}
   \NP {\bf B377} (1992) 143}

\REF\dasjevicki{S.~R.~Das and A.~Jevicki,
   \art{String field theory and physical interpretation of $D=1$ strings,}
   {\it Mod.\ Phys.\ Lett.\/} {\bf A5} (1990) 1639}
\REF\polchinski{J.~Polchinski,
   \art{Classical limit of 1+1 dimensional string theory,}
   \NP {\bf B362} (1991) 125}
\REF\grossklebanov{D.J.~Gross and I.~R.~Klebanov,
   \art{$S=1$ for $c=1$,}
   \NP {\bf B359} (1991) 3}
\REF\avanjevicki{J.~Avan and A.~Jevicki,
   \art{Classical integrability and higher symmetries of collective
   string field theory,}
   \PL {\bf B266} (1991) 35}
\REF\mooreplesser{G.~Moore and R.~Plesser,
   \art{Classical scattering in 1+1 dimensional string theory, hep-th/9203060,}
   \PR {\bf D46} (1992) 1730}

\REF\barton{G.~Barton,
   \art{Quantum mechanics of the inverted oscillator potential,}
   \AP {\bf 166} (1986) 322}
\REF\funk{M.~Carreau, E.~Farhi, S.~Gutman, and P.F.~Mende,
   \art{The functional integral for quantum systems with hamiltonians
   unbounded from below,}
   \AP {\bf 204} (1990) 186}
\REF\moore{G.~Moore,
   \art{Double-scaled field theory at $c=1$,}
   \NP {\bf B368} (1992) 557}

\REF\francis{F.E.~Low and P.F.~Mende,
   \art{A note on the tunneling time problem,}
   \AP {\bf 210} (1991) 380}

\REF\korepin{V.E.~Korepin,
   \art{Above-barrier soliton reflection,}
   {\it Pis'ma Zh.\ Eksp.\ Teor.\ Fiz.\/} {\bf 23} (1976) 224}
\REF\jackiwwoo{R.~Jackiw and G.~Woo,
   \art{Semiclassical scattering of quantized nonlinear waves,}
   \PR {\bf D12} (1975) 1643}

\REF\kresh{K.~Demeterfi, A.~Jevicki and J.P.~Rodrigues,
   \art{Scattering amplitudes and loop corrections in collective field
   string theory,}
   \NP {\bf B362} (1991) 173}
\REF\MSW{G.~Mandal, A.M.~Sengupta and S.R.~Wadia,
   \art{Interactions and scattering in $d=1$ string theory,}
   \MPL {\bf A6} (1991) 1465}

\REF\brusteinovrut{R.~Brustein and B.A.~Ovrut,
   \art{Stringy instantons,}
   U.~Penn. preprint UPR-522T, hep-th/9209045;
   \art{Non-perturbative interactions in string theory,}
   preprint UPR-524T, hep-th/9209033;
   \art{Non-perturbative effects in 2-D string theory or beyond the
   Liouville wall,}
   preprint UPR-523T, hep-th/9209081 }

\REF\martinec{E.J.~Martinec and S.L.~Shatashvili,
   \art{Black hole physics and Liouville theory,}
   \NP {\bf B368} (1992) 338}

\REF\witten{E.~Witten,
   \art{On string theory and black holes,}
   \PR {\bf D44} (1991) 314}


\date={October 1992}
\pubtype={}
\Pubnum{\vbox{ \hbox{BROWN-HET-880} \hbox{hep-th/9211049} }}
\titlepage

\title{\seventeenbf
   {Semiclassical Tunneling in 1+1~Dimensional String Theory}
   \footnote{\star}{Research supported in part by
   the U.S.~Department of Energy under contract DE-AC02-76-ER03130.
   E-mail: {\tt lee@het.brown.edu, mende@het.brown.edu}
   }}

\author{Julian Lee and Paul F.~Mende}

\address{\scrunch
   Department of Physics 		\break
   Brown University			\break
   Providence, Rhode Island~~02912
   \smallskip}

\abstract{ \scrunch
   We describe time-dependent tunneling of massless particles in
   1+1 dimensional string field theory.  Polchinski's
   description of the classical solutions in terms of the Fermi sea
   is used to identify the leading instanton contribution connecting
   the two half-lines.
   The field theory lagrangian is proportional to $1/g^2$, where
   $g$ is the string coupling constant, but the
   $S$-matrix for tunneling from one half-line to the other
   behaves as $\exp(-C/g)$.
   We note the constant~$C$ involves curious boundary contributions
   and observe that a prescription connecting the two half-lines unifies
   treatments of the Fermi level above and below the barrier.
   \smallskip}

\vfil \eject


\Chapter{Introduction}

String theory promises to provide a consistent theory of
gravitational and quantum physics.
Two-dimensional string theories
(for a review see Ref.~\refmark{\klebanov})
provide solvable models in which many of
the essential properties can be explored.
Among them are non-perturbative effects, spacetime singularities
and horizons, time-dependent backgrounds, background independence,
string symmetries, and string symmetry breaking.
Other effects, such as the full role of non-locality, are probably {\it not\/}
among them since in two dimensions, where the string has no
transverse oscillations, it resembles local point-particle field theory
as much as it can.
This is unfortunate since the extended nature of strings is likely
the key to a quantum theory of gravitation in higher dimensions.
Nevertheless there is a great deal to be learned and, as emphasized
by Shenker\refmark{\shenker} reason to believe that non-perturbative
effects such as tunneling may indeed share behavior characteristic of
higher dimensional theories.

In this paper we examine one of the intriguing aspects of one
of the richer such theories yet to be solved, the string field
theory for~$c=1$.
This theory perturbatively describes a massless scalar excitation
(possibly augmented by discrete contributions) living on one of two
half-lines.
We shall consider these to be weakly coupled through the repulsive potential
barrier which defines the theory and compute the amplitude for excitations
to tunnel from one half-line to the other.
This is done for the semiclassical limit of the collective field
Lagrangian by the standard method of solving the Euclidean time
field equations with the boundary conditions appropriate to tunneling.
This is all the more straightforward since Polchinski has given a
general solution for real-time evolution of the classical fields.

We show that tunneling amplitudes behave as $\exp(-C/g)$,
where~$C$ is a constant,
even though the Lagrangian is proportional to~$1/g^2$.
The reason for this is found to be a simple vestige of the underlying fermions
which have all but disappeared in the passage to the collective field.
We give an expression for~$C$ in terms of an arbitrary initial field
configuration.
Curiously $C$ has both a local contribution, largely independent of
the details of the initial field, and a non-local contribution.
The latter piece arises from boundary terms in the Lagrangian which
mix left- and right-movers.
These terms are absent in the Hamiltonian.
We also give examples of algebraic solutions to the field equations, which
might be explicit enough to help unravel the role of the boundary effects,
although we do not study this question here.

Our analysis is strictly limited to the tunneling effects of the field theory
defined below in Eq.~(2.1).
This is in contrast, for example, with the solitons discussed
in Ref.~\refmark{\jevicki}; in that paper the effective Lagrangian had no
repulsive potential but did have additional induced interactions from higher
orders in the $1/N$ expansion.
In terms of matrix models, tunneling as a source of non-perturbative unitarity
for half-line physics has been discussed in great detail
in Ref.~\refmark{\MPR}.

\Chapter{Solving the classical string field theory}

The collective string field theory\refmark{\dasjevicki}
gives perhaps the clearest description of the non-critical string theory
in $1+1$ Minkowski spacetime. It is described by the Lagrangian
$$
   L = {1\over g^2}\int\D x\,\left\{
   {1\over 2}{(\PHIDOT)^2\over\PHI} - {\pi^2\over 6}(\PHI)^3
   + {x^2 - 1\over 2}\PHI
   \right\}
   .\eqn\cft
$$
for the scalar field~$\zeta(t,x)$.

The principal motivation for studying low-dimensional field theories is to
uncover their non-perturbative dynamics, yet this is
difficult to achieve in the collective field formulation because the
field equations are non-linear and non-local.
The time-independent solution, the ``static tachyon background,''
$$
   \PHI = \cases{ {1\over \pi} \sqrt{x^2-1}, & \quad $|x|\ge 1$ \cr
   0, & \quad $|x| < 1$ \cr}
   \eqn\static
$$
was discovered immediately but until recently the
time-dependent problem has remained elusive.

\section{The Fermi sea}

Polchinski has clarified the problem of finding classical
solutions\refmark{\polchinski},
and in this paper we generalize his solution to include quantum tunneling.
In the free-fermion formulation of the $c=1$ model, the
static solution~$\static$
corresponds to filling states of the Fermi sea on both sides of the
potential $V(x)=-{1\over 2}(x^2-1)$.
Polchinski noted that time-dependent solutions of the collective
field theory~$\cft$ can be described by a dynamical Fermi surface:
 by deforming the static surface {\it in phase space\/}
($p^2=x^2-1$),
it evolves with time according to the equations of motion
for a collection of particles in an inverted oscillator potential.
The upper and lower edges of the Fermi surface, $p_{\pm}(t,x)$,
are precisely the left- and right-moving
collective fields\refmark{\polchinski,\grossklebanov}
(and denoted~$\alpha_\pm$ in Ref.~\refmark{\avanjevicki})
$$
   p_\pm = - g^2 \PI \pm \pi\PHI,
   \eqn\pdefs
$$
or
$$
   \PHI = {p_+ - p_- \over 2\pi}, \qquad
   \PI  =-{p_+ + p_- \over 2 g^2}.
   \eqn\xidefs
$$

Once specified at an initial time, $t=0$, the evolution of the surface
is given essentially by introducing action-angle variables.
Letting~$\sigma$ parameterize points along the surface, it is
$$
   \eqalign{
   x(t,\sigma) &= -a(\sigma)\, \cosh(t-\sigma), \cr
   p(t,\sigma) &= -a(\sigma)\, \sinh(t-\sigma). \cr
   }\eqn\evolution
$$

The static solution Eq.~$\static$ is given by $a(\sigma) \equiv 1$.
For arbitrary functions~$a(\sigma)$
these equations describe the evolution of the surface on the left-hand
side of the potential, $x<0$.  There is of course a corresponding
(independent) evolution on the right-hand side in terms of an
independent profile~$\tilde a(\sigma)$.
We focus first on the left-hand side, where disturbances
originate, and set $\tilde a(\sigma)\equiv 1$.

One can make contact with the collective fields by
eliminating~$\sigma$.
Inverting~$x(t,\sigma)$ gives $\sigma(x,t)$, where for fixed time
there are two solutions:
at each $x$ there is a point~$\sigma_+$ on the
upper surface and a point $\sigma_-$ on the lower surface.
Inserting into $p(t,\sigma)$, these solutions may be called~$p_\pm(t,x)$.
The time evolution is given by the equation
$$
   \partial_t p_\pm = x - p_\pm \partial_x p_\pm
   = \partial_x\left( {x^2-p^2\over 2} \right)
   .\eqn\eom
$$

\section{Non-parametric solutions}

Alternatively, $\sigma$ can be eliminated directly:  since
$$
   \eqalign{
   \sigma &= t - \cosh^{-1}\left(x/a(\sigma)\right) , \cr
   x^2 &- p^2 = a^2(\sigma) , \cr
   }\eqn\?
$$
these combine to give a functional, rather than a differential,
equation for~$p(t, x)$:
$$
   x^2 - p^2 = a^2\left(t - \cosh^{-1}\left({x\over \sqrt{x^2-p^2}}\right)
   \right).\eqn\?
$$
For some profiles~$a(\sigma)$ the equation for the Fermi surface
even becomes algebraic.  For example, if
$$
   a^2(\sigma) \equiv 1 - {\eta_0\over \cosh^n\sigma}
   ,\eqn\etadef
$$
where~$\eta_0$ is a constant, one can obtain polynomial solutions.
If $n=1$,
we obtain a sixth degree polynomial to solve for the time-dependent
classical solution(s)~$p(t, x)$:
$$
   \left(p^2 - x^2 + 1\right)^2\,
   \left(p\sinh t \pm x \cosh t\right)^2
   = -\eta_0^2 \left(p^2 - x^2\right)
   .\eqn\POLY
$$

\Chapter{Small fluctuations}

Now we concentrate on the propagation of small disturbances to the
static background.
This is a natural starting point for examining
particle-like propagation, scattering,  and later
quantization.
Moreover, small bumps tunnel most easily to the ``other world,''
the right half-line $x>0$, as will be apparent.

First is must be noted that the disturbance has to be small enough
that it does not ``double up,'' creating a fold in the Fermi
surface.
A sufficient condition is that the surface have
precisely one vertical asymptote in phase space at fixed~$t$.
Thus
$$
   0 = {\D x\over\D p}\Bigg\vert_t
   = {\partial x/\partial\sigma \over \partial p/\partial\sigma}\Bigg\vert_t
   = {a\sinh(t-\sigma)-a'\cosh(t-\sigma) \over
   a\cosh(t-\sigma)-a'\sinh(t-\sigma)}
   \eqn\?
$$
has one solution.
As the denominator is finite, there is exactly one solution to
$$
   a'/a = \tanh(t-\sigma) \approx a'
   .\eqn\?
$$
For a single-peaked~$a(\sigma)$ this is guaranteed if $a'(\sigma)\ll 1$,
or for the example of Eq.~$\etadef$, $\eta_0\ll 1$.
It is natural to identify the size of the
perturbation, $\eta_0$, with the string coupling~$g$ and take the
limit~$g\to 0$.
Next it should be noted that two constraints of the matrix model enter
naturally.
The first, that the eigenvalue density be positive,
$\PHI \ge 0$ is guaranteed by the above, since $p_+\ge p_-$.
The second, the normalization condition on the tachyon field, is
automatic for fluctuations about a normalized static solution.

Polchinski analyzed and interpreted the propagation of such
disturbances in Ref.~\refmark{\polchinski}.
By decomposing into modes of the tachyon field the asymptotic outgoing
wave, one can read off the leading terms of the tree-level $S$-matrix.
Moore and Plesser\refmark{\mooreplesser} extended this
analysis to all orders in the number of particles.

The picture that emerges is this:
semiclassical
particle states exist and are confined to the half-line~$x<0$, with
particle production occurring ``at the wall.''
This boundary interaction completes
the result of Gross and Klebanov\refmark{\grossklebanov} that
there is no scattering in the bulk:  $S=1$ away from the ``wall.''

\Chapter{Tunneling}

\section{General remarks}

But what of the other half-line, $x>0$?
Half-line physics recalls a number of physical contexts.
Reduction of spherically symmetric dynamics to radial coordinates
naturally defines one-dimensional physics in a coordinate $r\ge 0$;
the half line arises as an orbifold since $r$ and $-r$ are identified.
An impenetrable, repulsive interaction can dynamically confine particles
to a region of space.  In this case, there may be a non-trivial family
of boundary conditions that can be imposed, consistent with total
reflection.  Lastly, there may be a strong repulsive potential
which nearly confines, making the half-line a good approximation
and perturbative starting-point.
In all cases, physical considerations must determine the appropriate
boundary conditions on the wavefunctions and/or fields, and whether
propagation the right half-line is allowed or meaningful.
Here we bypass the interesting possibility of regarding $x$ as a radial
coordinate and view~$x$ as ranging over both half-lines.

Clearly from the quantum mechanics of the $-x^2$
potential\refmark{\barton, \funk, \moore},
fermions must be able to penetrate the barrier.
The possible leakage of particles through the potential has of
course been noted before.
Our goal is to use the free-fermion picture
to identify the origin of
the tunneling and then to use the collective field theory to compute
the tunneling amplitudes and leading behavior of the $S$-matrix
from the Euclidean time continuation of Eqs.~$\evolution$
which give the dominant field configuration.

The dynamical surface encloses an infinite number of fermions, so
it is helpful to recall first what happens to a single
particle of negative energy
incident from $x=-\infty$ on a generic repulsive potential.
Classically, the particle follows a  trajectory
$(x(t), p(t))$
in phase space until reaching the classical turning point (-a, 0)
before returning to $x=-\infty$ along the branch with $p<0$.

Quantum mechanically, this process is described by a wave packet,
suitably localized in energy and in position space, so that as
usual its peak follows the classical trajectory described above.
The reflected wave corresponds to the $p<0, x<0$ branch of
the phase space curve.

In general, part of the wave packet is transmitted --- it tunnels through
the barrier.
This part of the wave function can also be described in phase
space by noting that after the particle arrives at
$(-a,p)$ on the left
it instantaneously jumps to $(+a, p)$
on the right and continues out
along the $p>0, x>0$ branch ({\it cf.} Ref.~\refmark{\francis}).
That is, the transmitted packet emerges at the right-hand turning point
at $t=0$, just as the incident packet hits, and its peak follows the
classical trajectory along the right-hand branch.

How does the jump take place?
In phase space the particle simply continues along its classical curve
$p^2 = 2(E-V(x))$
where in order to get from one classical turning point ($x=-a$)
to the other ($x=+a$)
the momentum~$p$ becomes imaginary.
The tunneling factor of the time-independent wave function is
given by integrating $\exp(i\int p\,\D x)$ along this trajectory.
One can also describe
the tunneling in time-dependent language by saying that the particle evolves
in Euclidean time between the turning points.

Now we see how to generalize Polchinski's time-dependent
scattering-type solutions to tunneling.
The state is a collection of free particles inside the Fermi surface,
each moving along a classical trajectory.
If these particles tunnel --- time becoming Euclidean ---
eqs.~$\evolution$
 still describes a classical solution and stationary point of the action;
and the solution of highest symmetry is the minimum of the action.

\section{Evaluating the action}

To describe this instanton in more detail,
we introduce a symmetric disturbance, a surface
profile $a(\sigma) = a(-\sigma)$ where $1-a$ is peaked around $\sigma=0$
and is positive.
This is a bump on the static tachyon background.
It evolves
according to Eq.~$\evolution$ until it hits
the classical turning point at time
$t=0$, whereupon time is rotated, $t\to -i\tau$, and
it evolves in Euclidean time
until the disturbance emerges on the right hand side, and thence out to
$x=+\infty$.
Leaving aside heuristics and single quantum particles,
we return to our field theory.

The action for such a field configuration is
easily computed.
$p_\pm(t, x)$ should now be viewed as a quantum
field in 1+1 dimensions where
$$
   I = \int\D t\,\left( \int\D x\,\PI\PHIDOT - H \right)
   ,\eqn\?
$$
and where $H$ is the Hamiltonian of the system,
$$
   \eqalign{
   H = E &= {1\over 2\pi g^2} \int\D x\D p\, {1\over 2}
   \left(p^2 - x^2 +1\right)
   \cr &
   ={1\over 2\pi g^2} \int\D x \left(
   \left({p_+^3\over 6} - {(x^2-1)p_+\over 2}\right)
   - \left({p_-^3\over 6} - {(x^2-1)p_-\over 2}\right)\right)
   .}\eqn\?
$$
Since we are
only interested in having the disturbance tunnel (and not the entire sea)
the action of the static solution is to be subtracted off before doing
the rotation.

The Hamiltonian separates into left- and
right-movers\refmark{\grossklebanov,\polchinski,\avanjevicki}.
The full Lagrangian may be separated as well
using the equations of motion,
at the apparent cost of introducing non-locality.
One finds using $\xidefs$ that
$$
   \int\D t\D x\,\PI\PHIDOT = -{1\over 4\pi g^2}\int\D t\D x\,
   \left( p_+\partial_x^{-1} \dot p_+  - p_-\partial_x^{-1} \dot p_-  \right)
	+ B
   ,\eqn\legendre
$$
where $B$ contains cross terms mixing left- and right-movers.
It vanishes under naive integration by parts, but there are boundary
pieces to be considered below.
Using the equations of motion $\eom$,
$$
   \partial_x^{-1} \dot p_\pm = {1\over 2}(x^2 - p_\pm^2).
   \eqn\?
$$
the non-local operator~$\partial_x^{-1}$ is eliminated and the action
becomes
$$
   I = {1\over 2\pi g^2}\int\D t\D x\, \left(
   \left({p_+^3\over 12} + {x^2p_+\over 4} - {p_+\over 2} \right)
   - \left( p_+ \to p_- \right) \right)
   .\eqn\?
$$

It is easier to change variables and perform the integrals
over $\sigma$, $\sigma\in [-\infty,\infty]$,  instead of over $x$,
and transforming to fields $p(t,\sigma)$.
$$
   I = {1\over 2\pi g^2}\int\D t \int_{-\infty}^\infty\D \sigma\,
   \left( {\partial x\over \partial\sigma} \right)_t
   \left({p^3\over 12} + {x^2p\over 4} - {p\over 2} \right)
   .\eqn\theaction
$$

Regarding this as a two-dimensional field theory, we obtain the
amplitude by integrating the action over all field configurations
with appropriate initial and final states\refmark{\korepin,\jackiwwoo}:
$$
   \int{\cal D}\{\PHI\} {\rm e}^{i I}
   .\eqn\?
$$

Having identified the saddle point, we simply extract the leading
exponential dependence, leaving questions of phases, fluctuations,
and subtleties regarding the measure for future work.
Hence we integrate only the part of the action in the classically
forbidden region, interpolating between the left- and right-hand
solutions, \ie, from $\tau\equiv it=0$ to $\tau=\pi$.

Therefore we insert the field configuration~$\evolution$ into the action
eq.~$\theaction$ and integrate over Euclidean time to obtain
the result
$$
   I = {i \over 8g^2} \int_{-\infty}^\infty\D \sigma\,
   \left( 1 -  a^2(\sigma) \right)
   .\eqn\Action
$$
The unperturbed action has been subtracted:  $I$ vanishes if $a=1$.

\section{One lump or two?}
Hence this contributes to the tunneling amplitudes as
$$
   \A \sim {\rm e}^{i I} = {\rm e}^{-|I|}
   .\eqn\?
$$

This is quite general and hold for a large class of profiles $a(\sigma)$.
How do we estimate the importance of this process?

The tunneling is clearly largest when the perturbation is smallest.
But $1-a$ cannot be made arbitrarily small.
The continuum picture of the Fermi liquid in phase space is valid as
a limiting case as $g\sim\hbar\to 0$.  For $g\equiv 0$ there is obviously
no tunneling, or other interaction, so it must be kept finite.

The action~$\Action$ has a simple interpretation:
it is proportional to the phase space volume of the perturbation.
Indeed, since the volume of the sea is
$$
   V = \int \D x\, \D p
   = {1\over 2}\int\D\sigma\,
   \left( x {\partial p\over \partial\sigma}
   - p {\partial x\over \partial\sigma} \right)
   ,\eqn\?
$$
the volume of the perturbation is given by subtracting the (divergent)
non-interacting sea,
$$
   V_{\rm pert}
   = {1\over 2}\int\D\sigma\, \left( 1 - a^2(\sigma) \right)
   ,\eqn\?
$$
so that
$$
   \A \sim  {\rm e}^{-V_{\rm pert}/4g^2}
   .\eqn\?
$$

Thus for finite $g\sim\hbar$, a bump on the Fermi surface must have
volume at least
$$
   V_{\rm pert} = 2\pi\hbar = 2\pi g
   ,\eqn\?
$$
the phase space
volume for adding an eigenvalue to the left hand side of the
well.
Actually, this is the volume for a one non-relativistic fermion,
and a pair is required to make an asymptotic boson state.
Therefore the $g$-dependence of the largest non-perturbative tunneling is
$$
   \A \sim {\rm e}^{-\pi/g}
   .\eqn\?
$$
This result --- that tunneling processes in string field theory behave
as $\Exp{-{\rm const.}/g}$ as opposed to $\Exp{-{\rm const.}/g^2}$
in field theory --- is in agreement with the arguments and calculations
emphasized by Shenker\refmark{\shenker}.

\section{Boundary terms}

Now we consider the term~$B$ in Eq.~$\legendre$.
$$
   B = {1\over 4\pi g^2}\int\D t\D x\,
   \left( p_+\partial_x^{-1} \dot p_-  - p_+\partial_x^{-1} \dot p_-  \right)
   .\eqn\bdy
$$
This boundary term is the only part of the action which mixes
left and right movers, \ie, which couples the fields $p_+$ to $p_-$
and it must be analyzed carefully.
Using again the equations of motion Eq.~$\eom$,
$p=\partial x/\partial t|_\sigma$,  and changing integration
variables,
$$
   \eqalign{
   B &= {1\over 8\pi g^2}\int\D t\D x\,
   \left(p(t,\sigma_+) a^2(\sigma_-) - (p(t,\sigma_-) a^2(\sigma_+)\right)
   \cr &
    = {-1\over 8\pi g^2}\int\D t\int_t^\infty\D \sigma_+\,
   \left({\partial x\over \partial \sigma_+}\right)_t
   \left({\partial x\over \partial t}\right)_\sigma a^2(\sigma_-)
   \cr & \qquad
    - {1\over 8\pi g^2}\int\D t\int_{-\infty}^t\D \sigma_-\,
   \left({\partial x\over \partial \sigma_-}\right)_t
   \left({\partial x\over \partial t}\right)_\sigma a^2(\sigma_+)
   .}\eqn\?
$$
These combine to a single integral if we define a function $\Sigma$
for the non-local behavior:
$$
   \eqalign{
   \Sigma(t,\sigma_+) &= \sigma_-, \qquad \sigma_+ > t,
   \cr
   \Sigma(t,\sigma_-) &= \sigma_+, \qquad \sigma_- < t
   .}\eqn\Sigmadef
$$
Integrating by parts with respect to time, one obtains
$$
   B = -{1\over 16\pi g^2}\int\D t\D\sigma\,
   \left[ a^2(\sigma)\,a^2(\Sigma(t,\sigma)) -1 \right]
   .\eqn\?
$$

\Chapter{Discussion}
\section{Results}

Combining with the previous result for the local terms, we find
$$
   \eqalign{
   \A \sim \exp(i I) &= \exp\Bigl(
   -{1\over 8g^2}\int\D\sigma (1-a^2(\sigma))
   \cr & \qquad\qquad
   + {1\over 16\pi g^2}\int_0^\pi\D\tau\int\D\sigma (1-a^2(\sigma)a^2(\Sigma))
   \Bigr)
   \cr &= \exp\Bigl(-{\pi\over g}
   + {1\over 16\pi g^2}\int_0^\pi\D\tau\int\D\sigma (1-a^2(\sigma)a^2(\Sigma))
   \Bigr)
   .}\eqn\?
$$

This behaves generically
as $\exp(-C/g)$, where $C$ is constant, although
without a definite choice of $a(\sigma)$ it is difficult to analyze
$\Sigma$ and evaluate the exact coefficient~$C$.
By studying explicit examples, such as those Eq.~$\POLY$, one might
be able to learn more.

This result is in contrast with the typical behavior of field theories,
in which non-perturbative contributions enter as
$\exp(-{\rm const.}/g^2)$ ({\it cf.\/}~\refmark{\shenker}).
While the former type of contribution is formally larger in the
weak-coupling limit, one cannot evaluate the actual size of non-perturbative
effects without knowing the constants, prefactors, \etc, not to mention
the physics which determines the actual coupling~$g$.
One may conclude that the full string theory does contain non-perturbative
effects other than those contained in the low-energy effective field
theory limit of the string; this should not be surprising.

It is clear that this is the continuum collective
field analog of the single eigenvalue instanton.
We stress that the relevant field configurations are everywhere smooth, even
at the turning points.
One apparent source of singularities in analyzing the collective field comes
about if one chooses to expand the field around the static solution
$\phi = \phi_0 + \partial_x\xi$.
In contrast, we have not expanded perturbatively:
we computed the full action of background+fluctuation relative to the
static action.

{\section{Tunneling and the half-line scattering amplitudes}}

As an additional application of tunneling behavior, we show that
there is a natural prescription to connect the
theories defined with the Fermi level either above or below the barrier.
This completes the work of Demeterfi, Jevicki and
Rodrigues\refmark{\kresh}\ and
makes contact with that of Klebanov\refmark{\klebanov}.
(This is related to the prescription of
Mandal, Sengupta and Wadia\refmark{\MSW} as we shall see.
\foot{We thank G.~Moore for a discussion on this point.})
In these papers, string scattering amplitudes were computed from
the collective field Hamiltonian and found to agree.
The cubic interaction term is
$$
   \eqalign{
   H_3 &= {1\over 4\sqrt{2\pi^3}}{1\over 3L}\sum_{m,n,p}^\infty
   \sqrt{\omega_m\omega_n\omega_p}\, f_{m+n+p}\, :a_m a_n a_p:
   \cr & \quad +
   {1\over 4\sqrt{2\pi^3}}{1\over L}\sum_{m,n,p}^\infty
   \sqrt{\omega_m\omega_n\omega_p}\, f_{m+n-p}\, :a^\dagger_m a_n a_p:
   \,+\,{ \rm h.c.}
   }\eqn\?
$$
where $a_m$ are the standard operators of frequencies $\omega_m$
and $f_k$ are the three-string vertices to be discussed shortly.
The four-tachyon amplitude result is
$$
   T = -{i\over 16\pi} \sqrt{k_1 k_2 k_1' k_2'}
   \left[ |k_1+k_2| + |k_1-k_1'| + |k_1-k_2'| -4i \right]
   .\eqn\?
$$
One of the mysteries of this formula is whether the last, inhomogeneous
term which arises as an infinite sum of discrete contributions might
be related to tunneling behavior and resonant states under the barrier.

In Refs.~[\kresh,\MSW], the calculation was done with $\mu>0$, so that the
Fermi level was below the barrier height; and in Ref.~[\klebanov] it was
done taking $\mu<0$.
\foot{In the earlier sections $\mu>0$ was held fixed and scaled out
of the Lagrangian.}

For $\mu<0$, the calculation is very simple since the wave functions
are all decomposable into plane waves.
On the other hand, the spacetime interpretation is obscured
since the volume of space, to be identified with $\log\mu$,
is complex for $\mu<0$.
For $\mu>0$, however, one must deal with the turning point.
Clearly the amplitudes should be related through analyticity in $\mu$.

The form factors, or three string vertex functions, are given by integrals
of the form
$$
   f_k = \int {\D\tau \over \phi_0^2(\tau)}\theta_k(\tau)
   ,\eqn\fkdef
$$
where $\theta_k$ are the basis of wavefunctions appropriate to the
interval where the string field is defined.
To see how the form factors in the two definitions of the theory
are related, it is sufficient to consider the prototype integral
of the form
$$
   I(\mu) = \int{\D x \over \sqrt{(x^2-2\mu)^3}}
   ,\eqn\FORM
$$
where we drop the wavefunction $\theta_k$ and change variables from $\tau$
to $x$
\foot{It bears emphasizing that all calculations, here and in section~4,
may be done directly in position space without recourse to the
Euclidean time that serves to parameterize the relevant contours.}:
$$
   \tau = {1\over\pi}\int{\D x\over\phi_0(x)}
   .\eqn\?
$$

For $\mu>0$ and real, the physical case discussed at length in Ref.~[\kresh],
this type of integral was evaluated by cutting off below the classical
turning point at $x=-\sqrt{2\mu}$:
$$
   I_\epsilon(\mu) \equiv \int_{-\infty}^{-\sqrt{2\mu}+\epsilon}
   {\D x\over\sqrt{(x^2-2\mu)^3}}
   = -{1\over 2\mu} + {1\over (2\mu)^{3/4}\sqrt{\epsilon}}
   .\eqn\DJR
$$
These authors then dropped the second term, arguing that it was an irrelevant
non-universal term by first letting $\mu\to 0$ before removing the
cutoff, $\epsilon\to 0$.
They proceeded to consistently drop such turning point contributions.
One could argue that their results only hold strictly at $\mu=0$.
But it is easy to see how the same results hold for {\it finite\/}~$\mu$
if one takes the integrand to be analytically continued past the
turning point in the spirit of the previous section.
(Recall that $p = \pi\phi_0 = \pi\PHI$.)

For $\mu<0$ as in Ref.~[\klebanov], or more generally for any complex
$\mu$ off the positive real axis, the integral Eq.~$\FORM$
taken over the whole real axis is elementary:
$$
   I(\mu) = -{1\over \mu}, \qquad \mu\notin [0, \infty)
   .\eqn\?
$$
Taking the limit $\mu \to \mu_0 - i\epsilon$, $\mu_0\in[0,\infty)$,
and deforming the contour
slightly to avoid the turning points at $x=\pm(\sqrt{2\mu_0}-i\epsilon)$,
one sees that by continuing through the classically forbidden region,
rather than imposing a fixed cutoff as in Eq.~$\DJR$, one obtains
an $\epsilon$-independent result for any finite~$\mu>0$, not only
the extreme limit $\mu=0$.
This same contour prescription works for all the form factors~$\fkdef$
as well, since the wave functions introduce no new finite singularities
\foot{And taking appropriate care of course in closing the contours.}

In Ref.~\refmark{\MSW} a different regularizing contour was proposed
which circled the left-hand turning point in the complex $x$-plane and
returned to $x=-\infty$.
If one unfolds this contour at the turning point, swinging the return portion
by~$180^\circ$ so it runs along the whole real axis (yet avoiding the
right-hand turning point on the opposite side) one recovers the prescription
of the last paragraph, revealing the implicit below-barrier contribution.
In $\tau$-space, this amounts to deforming the contour from the
real axis to agree with that used in section~4.  The analyticity of the
form factor integrals under these deformations is easily verified.

To summarize, a natural prescription for deriving finite $\mu$
scattering amplitudes is to keep $\mu>0$ so that space is real
and to include the classically forbidden region.
This implicit contribution of tunneling regulates the form factors
and explains the agreement with the calculations done formally
for $\mu<0$.
\foot{The question of interpreting $\mu<0$ has also been
raised\refmark{\martinec} in the context of black holes\refmark{\witten}.}

\section{Related work}

After this work was completed we received a series of papers
by Brustein and Ovrut also dealing with tunneling in the
collective field theory\refmark{\brusteinovrut}.
In contrast to the scattering considered here,
they compute the tunneling between static tachyon vacua.
They also relate their results to the effective Liouville theory
and write down operators induced by non-perturbative effects.
These authors stress computing first for finite~$N$
before taking various scaling limits inside and outside.
Here we have worked directly in the limiting classical collective field
theory.
The results, though similar, are not identical.
In the collective field theory at $N=\infty$, their field configuration
is singular, quite reminiscent of the single-eigenvalue instanton
of the matrix model; ours is smooth.
Both give similar contributions to the
action.

\ack
It is a pleasure to thank A.~Jevicki for many helpful discussions.

\refout
\end